\documentclass[preprint,aps,showkeys,preprintnumbers,onecolumn,showpacs,amsmath,amssymb,floatfix,endfloats*]{revtex4}
\usepackage{graphicx}% Include figure files
\usepackage{dcolumn}% Align table columns on decimal point
\usepackage{bm}% bold math
\begin{document}
\preprint{\textit{Version Feb 14th 08}}
\title{A Charge and Spin Readout Scheme For Single Self-Assembled Quantum Dots}
\author{D. Heiss}
\author{V. Jovanov}
\author{M. Bichler}
\author{G. Abstreiter}
\author{J. J. Finley}
\email{finley@wsi.tum.de}
\affiliation{Walter Schottky Institut, Technische Universit\"at M\"unchen, Am Coulombwall 3, D-85748 Garching, Germany}
\date{\today}
\begin{abstract}
We propose an all optical spin initialization and readout concept for single self assembled quantum dots and demonstrate its feasibility. Our approach is based on a gateable single dot photodiode structure that can be switched between charge and readout mode. After optical electron generation and storage, we propose to employ a spin-conditional absorption of a circularly polarized light pulse tuned to the single negatively charged exciton transition to convert the spin information of the resident electron to charge occupancy. Switching the device to the charge readout mode then allows us to probe the charge state of the quantum dot (1e, 2e) using non-resonant luminescence. The spin orientation of the resident electron is then reflected by the photoluminescence yield of doubly (X$^{2-}$) and singly (X$^{-1}$) charged transitions in the quantum dot. To verify the feasibility of this spin readout concept, we have applied time gated photoluminescence to confirm that selective optical charging and efficient non perturbative measurement of the charge state can be performed on the same dot. The results show that, by switching the electric field in the vicinity of the quantum dot, the charging rate can be switched between a regime of efficient electron generation ($\Gamma \gg $ 10$^{6} s^{-1}W^{-1}cm^2$) and a readout regime, where the charge occupancy and, therefore, the spin state of the dot can be tested via PL over millisecond timescales, without altering it. Our results show that such a quasi-continuous, non perturbative readout of the charge state of the dot allows to increase the dark time available for undisturbed spin manipulation and storage into the millisecond range, whilst still providing sufficient signal for high fidelity readout. Consequently, our readout scheme would allow the investigation of spin relaxation and decoherence mechanisms over the long timescales predicted by theory is possible. 
\end{abstract}
\pacs{	78.66.Fd, %III V low dimensional semiconductors optical properties
			 	78.67.De, %QD opt properties
}% PACS, the Physics and Astronomy
% Classification Scheme.
\keywords{Quantum Dots, GaAs, InGaAs, Readout, Charge, Spin}%Use showkeys class option if keyword display desired
\maketitle
The field of spin based quantum information processing using solid state nanostructures has made striding advances in recent years.\cite{Coish06} Perhaps the greatest progress has been made using electron spins trapped in  semiconductor quantum dots (QDs) \cite{Hanson07} or localized around isolated Nitrogen-Vacancy (NV) centers in diamond.\cite{Jelezko04a} Arbitrary single qubit rotations can now be performed using both systems \cite{Petta05,Koppens06,Nowack07,Jelezko04a} and, recently, a quantum register based on coupled electronic and nuclear spin qubits was demonstrated in the NV system.\cite{Jiang08} Whilst the spin coherence times are sufficiently long in both cases to allow for reliable state manipulation \cite{Petta05,Koppens06,Hanson06a}, a key step was the development of sensitive spin \emph{readout} methods, based on spin to charge \cite{Elzerman04} or spin to photon \cite{Gruber97,Jelezko04a,Hanson06b} conversion. Both approaches have their respective advantages and disadvantages: NV centers in diamond can be coherently manipulated at room temperature, whilst QDs typically require low temperatures due to their comparatively weak orbital quantization energies ($E_{e-QD}\sim0.1 - 50$meV$\equiv T_{e-QD}\sim 100$mK$ - 30$K). On the other hand, arrays of electrostatic \cite{Schroeer07} or self assembled QDs \cite{Xie95} can be readily formed making them potentially more scalable.\cite{Loss98} Unlike electrostatically defined QDs, self-assembled QD confine both electrons and holes simultaneously and are, thus, optically active. This allows the spin state of confined charges to be optically manipulated and interrogated. Both, electrostatically defined and self assembled QDs exhibit long electron spin coherence times $\sim$1~$\mu$s \cite{Hanson07,Greilich06}, limited only by coupling to the nuclear spin system.\cite{Tartakovskii07,Maletinsky07} However, all optical spin readout on single self assembled QDs is extremely challenging. Approaches based on spin to photon polarization conversion have been demonstrated both on ensembles \cite{Kroutvar04} and single dots \cite{Young07}. More recently, sensitive techniques based on direct absorption \cite{Hoegele05,Ramsay08} or time resolved Faraday \cite{Atatuere07} and Kerr \cite{Berezovsky06} rotation have been applied to test the spin of an isolated electron. Whilst all of these approaches work for single dots, the very weak signals involved inherently limit the time available for spin manipulation to $\sim$1~$\mu$s \cite{Young07} or even few nanoseconds \cite{Ramsay08, Berezovsky06}, whilst other experiments do not provide any time resolution \cite{Atatuere07,Hoegele05}. Even though spin manipulation can be performed very quickly using the AC Stark effect for example \cite{Ramsay08}, such approaches are not ideally suited to probe decoherence mechanisms that influence the electron spin over the long spin coherence times predicted by theory T$_{2} \gg $ 1 $\mu$s \cite{Golovach04} .  
\\
In this paper we propose a new scheme for optical spin readout in self-assembled InGaAs QDs and experimentally demonstrate its feasibility. The readout method provides the potential to extend spin storage times into the millisecond timescale, whilst keeping the experimental integration time manageable. Our approach is based on controlled optical charging of a single In$_{0.5}$Ga$_{0.5}$As dot incorporated within an electrically tunable Al$_x$Ga$_{(1-x)}$As-GaAs Schottky photodiode heterostructure. The read out concept employs the spin dependent absorption of a laser pulse tuned to the negatively charged exciton (X$^{-1}$) absorption line of a singly charged  (1e) QD, effectively converting the spin information into a more robust variable - the charge occupancy of the dot. Our measurements demonstrate that the devices can be switched between a regime where optical charging is highly efficient to one where optical readout using photoluminescence does not perturb the charge status of the dot. Consequently, charge and spin detection can be performed using time gated photoluminescence, as our results demonstrate, resulting in strong optical signals that considerably lengthen the time available to manipulate spins and study their relaxation and decoherence. 
\\
\begin{figure}[h]
	\centering
		\includegraphics[width=1.0\textwidth]{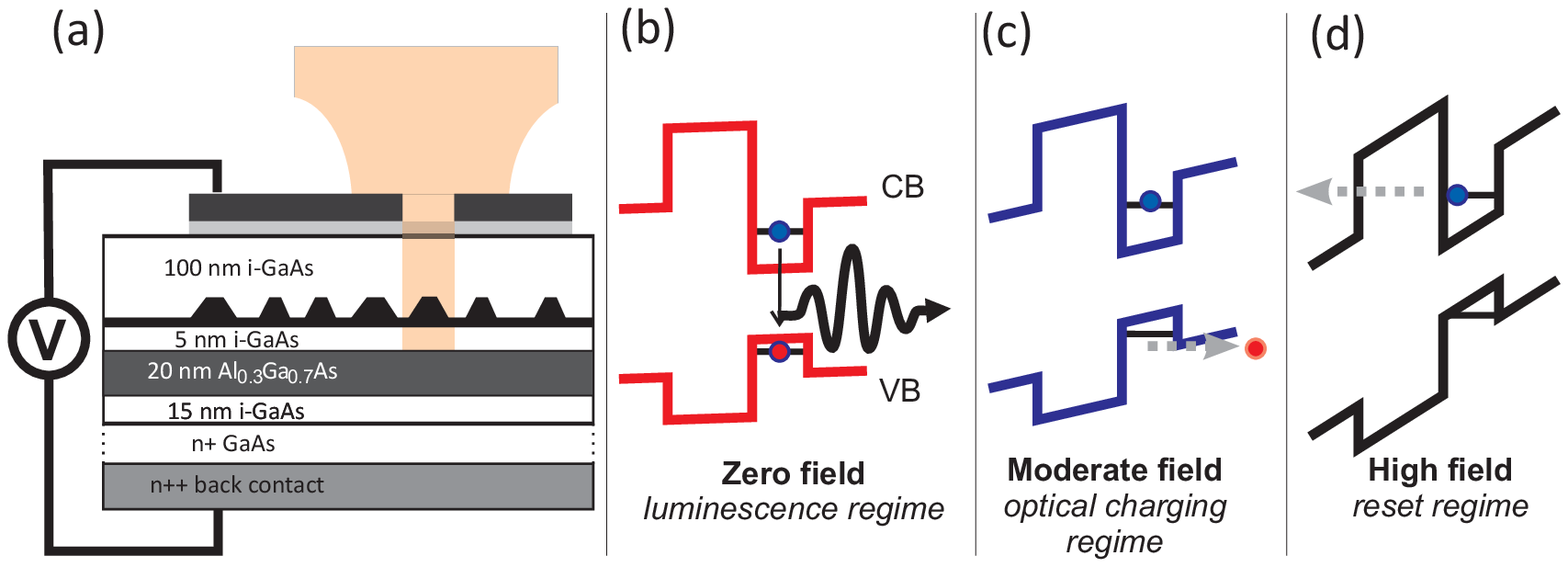}
	\caption{(Color online)(a) Schematic representation of the sample structure. We distinguish three electric field regimes:  (b) The luminescence regime at low electric fields. (c) Hole tunneling at moderate electric fields leads to the charging regime.(d) The reset regime, where electrons tunnel to the back contact. }
	\label{fig:Figure1}
\end{figure}
In order to perform optical charging experiments on a single self assembled QD precise control of the local electric field is necessary. This can be achieved by embedding the dot in the intrinsic region of a Schottky photodiode structure formed by a heavily n-doped back contact and a 5~nm thick, semi transparent Ti top contact. The epitaxial layer sequence of the devices investigated is depicted schematically in fig~1a. The intrinsic region has a total thickness of 140~nm with the QDs positioned 40 nm above the n-doped layer. This sample structure leads to a static electric field of 70~kV/cm per volt applied in the intrinsic region of the Schottky diode and a flat band condition ($|$F$|$=0~kV/cm) at an applied voltage of V$_{app}$=0.9~V. An opaque Au shadow mask is evaporated onto the sample surface to allow the optical selection of single dots through 1~$\mu$m diameter shadow mask apertures defined using polybeads spin coated onto the sample surface before metallization. To ensure that tunneling escape of holes from the dot is much faster than for electrons, thus, enabling optical charging, an asymmetric Al$_{0.3}$Ga$_{0.7}$As barrier with a width of 20~nm was grown below the QD layer. By controlling the voltage V$_{app}$ applied to the Schottky gate, the device can be operated in one of three regimes: (i) if the device is biased in forward direction so that the electric field across the QDs is small, both the tunneling times of electrons and holes are longer than the excitonic radiative lifetime ($\tau_{rad}$) and photoluminescence (fig~1b) will be observed. (ii) If the electric field is increased to the point where photogenerated holes tunnel out of the dot over timescales faster than $\tau_{rad}$, the dot charges up with electrons (fig~1c). (iii) If a large electric field is applied, electrons can tunnel through the AlGaAs barrier, hence, reducing the charge accumulation in the dot (fig~1d).
\\
To characterize these three distinct luminescence (i), optical charging (ii), and reset (iii) regimes we performed photo-luminescence measurements on a single QD as a function of the applied gate potential. These measurements were performed using a $\mu$PL-setup in a He bath cryostat at a temperature of 8~K. The sample was excited using a Ti-Sapphire-laser with a power density of 3~W/cm$^2$ tuned to the low energy side of the wetting layer absorption continuum ($\lambda_{exc}$=880~nm). The signal was detected using a 0.55~m single spectrometer and a LN$_2$ cooled Si charge coupled device multichannel detector (CCD). An aperture that showed only emission from a single QD was selected by carefully performing power dependent measurements. Typical $\mu$PL measurements recorded from this QD as a function of electric field are presented in the upper panel of fig~2. 
The most prominent emission lines arising from s-shell recombination in the dot can be observed in the range of 1339 - 1341~meV (fig~2a). As the electric field increases these emission lines shift parabolically towards lower energy due to the quantum confined Stark effect.\cite{Fry00} Furthermore, the intensity of the lines reduces at higher field due to carrier tunneling escape from the dot and emission is quenched entirely for $|$F$|>$ 45~kV/cm. Here, the tunneling time of the optically created holes becomes much faster than $\tau_{rad}$.
\\
\begin{figure}[h]
	\centering
		\includegraphics[width=1.0\textwidth]{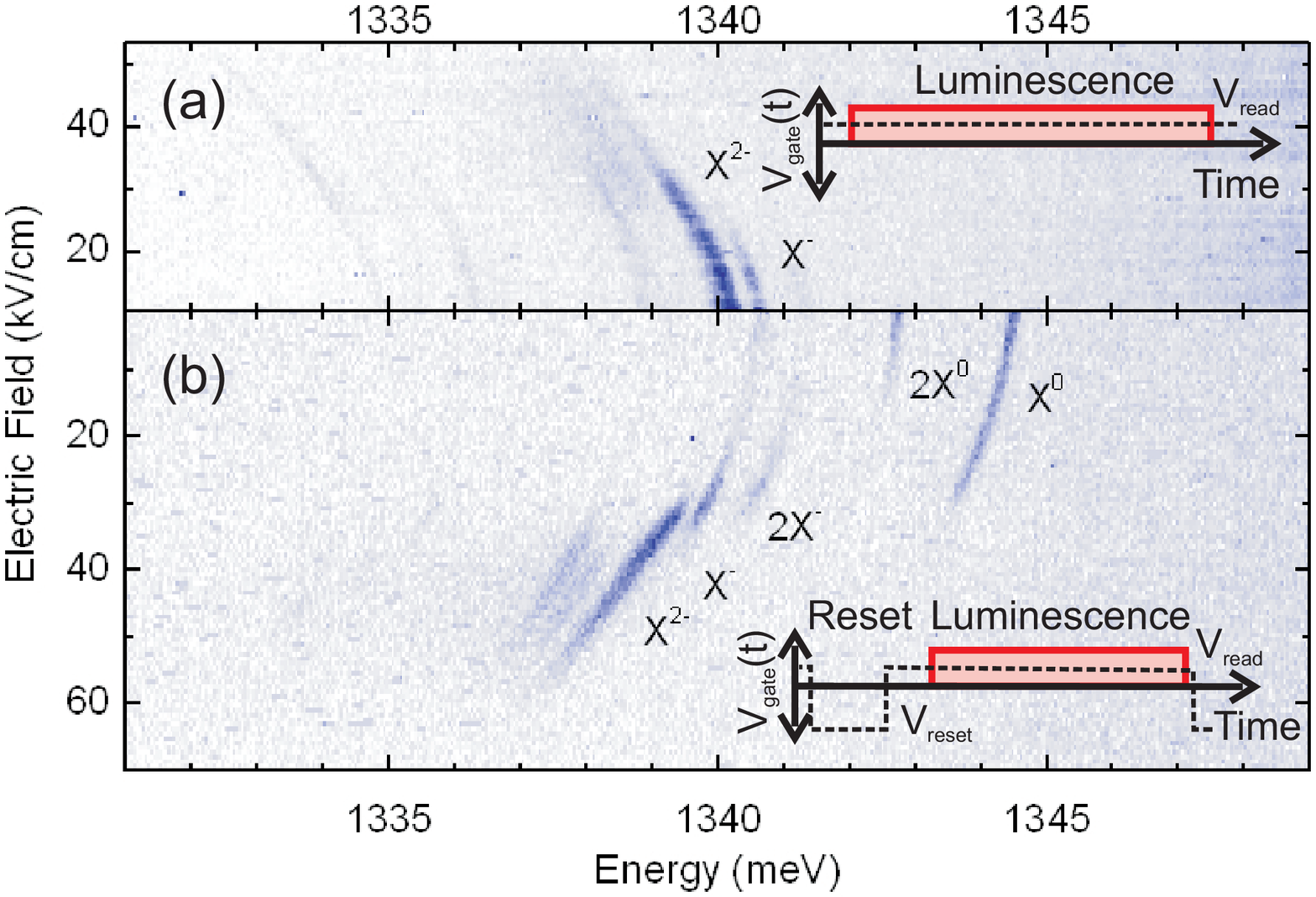}
	\caption{(Color online) Photoluminescence intensity for the neutral exciton (X$^0$) , negatively charged exciton (X$^{-1}$,X$^{-2}$)   and biexciton (2X$^{0}$,2X$^{-1}$) emission from a single QD as a function of emitted photon energy and applied electric field (a) with DC bias applied (b) with application of a reset pulse (see inset). With the introduction of a reset pulse charge accumulation is prevented and the charge neutral QD transitions become visible. }
	\label{fig:Figure2}
\end{figure}
From this measurement alone it is not possible to directly identify the excitonic states that give rise to the luminescence lines, since the asymmetric tunneling times of electrons and holes caused by the AlGaAs barrier may lead to an accumulation of optically generated electrons in the dot over the typical integration times of the experiment ($\sim$300~s).
To avoid electron accumulation in the dots, we replaced the DC-bias applied to the Schottky diode by the tailored voltage pulse sequence depicted schematically in the inset of fig~2b. The gate potential applied to the QD V$_{app}$(t) is periodically biased in reverse direction to allow electron tunneling and discharging of the QD. The waveform used to obtain the data presented in fig~2b consisted of a highly negative reset pulse (V$_{app}$(t) = V$_{reset}$=-2~V, $|$F$_{reset}| \sim$200~kV/cm) with a duration of 400~ns followed by 1.6~$\mu$s at a bias voltage V$_{app}$(t) = V$_{readout}$ (see fig~2b inset), where luminescence can be observed. The laser was turned ON 200~ns after switching the voltage from V$_{reset}$ to V$_{readout}$ for a readout time $\Delta t_{read}$=1$\mu$s to ensure illumination only when $V_{readout}$ is stable. The resulting PL spectra are presented in fig~2b as a function of $F_{readout}$. In contrast to fig~2a, the introduction of the reset pulse results in the emergence of two prominent emission lines, labeled 2X$^{0}$ and X$^{0}$ in fig~2b on the high energy side of the emission spectrum. These lines are visible for low electric fields, whilst the emission lines observed in fig~2a without the reset pulse are only observed in the high field region of the image plot. 
\\
The dramatic difference between the spectra recorded without and with the reset pulse can be readily explained by taking accumulation of optically generated charge in the QD into account. Due to the asymmetric AlGaAs barrier, the tunneling escape time for holes is much faster than for electrons. A fraction of excitons optically pumped into the dot do not, therefore, recombine radiatively, but instead result in charging of the dot with excess electrons. The hole tunnels towards the Schottky contact, while the electron stays trapped in the dot due to the presence of the AlGaAs barrier. Since the hole tunneling time is strongly field dependent, the rate at which this charging process takes place will increase at higher electric field applied to the dot. Nevertheless, even for small charging rates electrons will accumulate in the dot, if left unperturbed for the integration time of our experiment leading to luminescence of the negatively charged excitonic states X$^-$, 2X$^-$ and X$^{2-}$. In this case neutral excitonic states (X$^0$, 2X$^0$) cannot be observed, since the dot is charged with a very high probability. For higher order charge states one electron occupies the p-shell of the QD, where the electron tunneling time through the barrier is much shorter than the radiative lifetime of the exciton. As a result luminescence from highly charged states is weak and the charging process is self limiting. Further support to this assignment of the luminescence lines is given by a good agreement of the observed energy splittings in fig~2 with the typical renormalization energies observed in QDs \cite{Warburton00, Finley01}. Furthermore, the peak assignments presented in fig~2 are confirmed by the time evolution of the luminescence intensities of the charged states, discussed below.
\\
If, as described above, a reset pulse is introduced into each measurement cycle the electrons periodically tunnel out of the QD leaving it empty. In this case the emission from the uncharged states is restored, providing that the optical charging rate is much smaller than the repetition frequency \textit{f}$_{rep}$ of the reset pulse. If the charging rate exceeds \textit{f}$_{rep}$, as is the case for high electric fields in fig~2b, the dot will charge up and the charged exciton transitions again become visible. Thus, from the data presented in fig~2b we conclude that a illumination time of 1~$\mu$s is sufficient to charge the dot with 1 electron (2 electrons) on average at a field of $|$F$_{readout}|$=25~kV/cm (40kV/cm). This conclusion directly follows from the dominance of the X$^-$ (X$^{2-}$) spectral line in this range of electric field.
Closer inspection of the plot in fig~2b shows that the singly negatively charged lines (X$^-$, 2X$^-$) persist, even for very small electric fields. We attribute this observation to electrostatic charging from the n-doped back contact that plays a role at low electric fields.
Two of the three regimes of device operation described above can be directly identified in the plot in fig~2b. For fields $|$F$_{readout}| <$ 20~kV/cm optical charging of the QD does not occur within the readout time. In contrast, for fields $|$F$_{readout}|>$ 20~kV/cm optical charging takes place since the hole tunneling times become comparable to the radiative recombination lifetime of the exciton. Finally, for $|$F$_{readout}|>$55~kV/cm, fast hole tunneling prevents radiative recombination completely, such that luminescence cannot be detected. 
\\
\begin{figure}[h]
	\centering
	\includegraphics[width=1.0\textwidth]{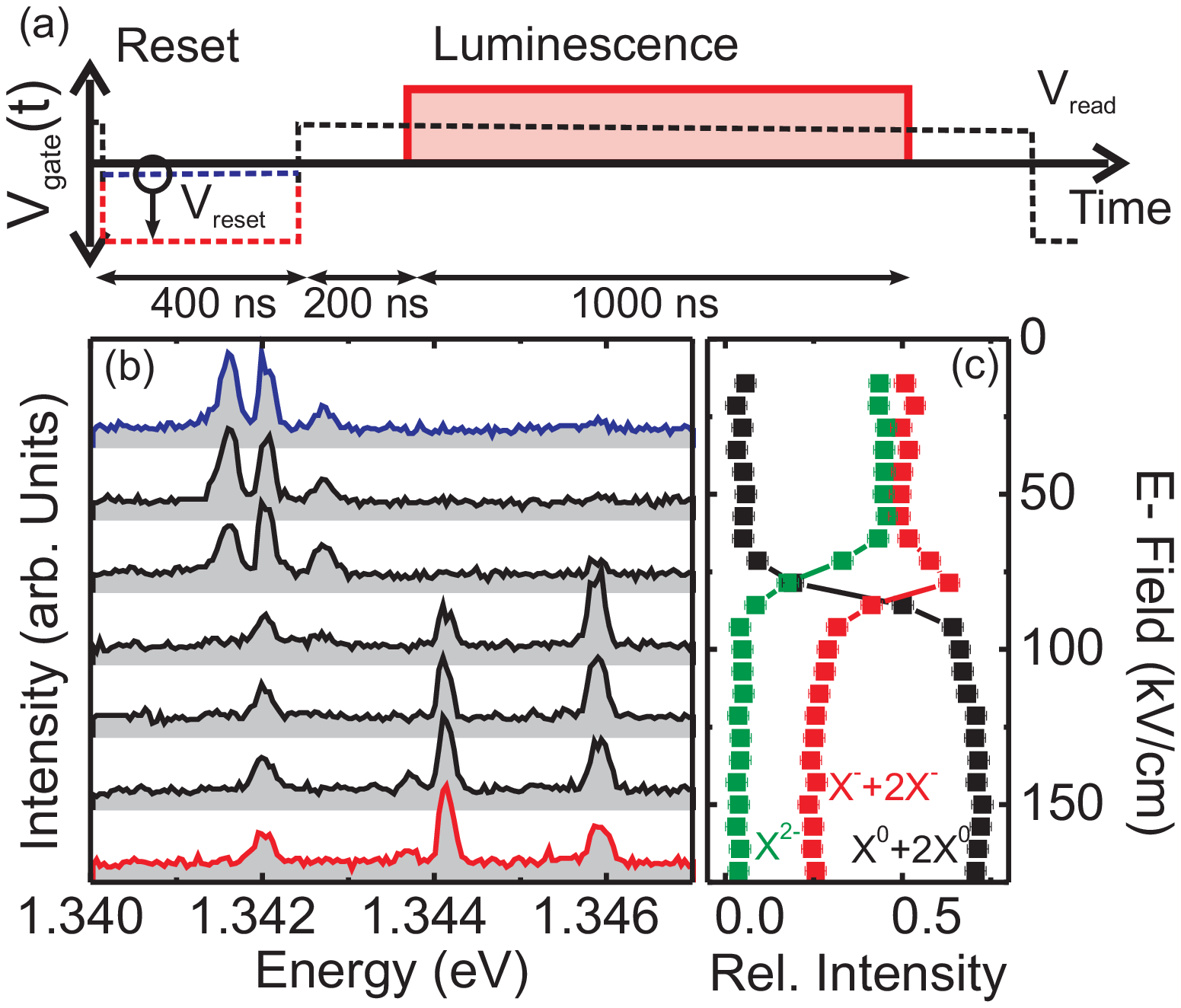}
	\caption{(Color online) (a) Schematic representation of the pulse sequence, V$_{reset}$ is changed during the experiment. (b) Photoluminescence spectra as a function of emitted photon energy for reset voltages ranging from V$_{reset}$=0.5V (blue) to V$_{reset}$=-1.5V (red) (c) Relative luminescence intensities of the charge neutral (X$^0$+2X$^0$, black), singly charged (X$^-$+2X$^-$, red) and the doubly charged (X$^{2-}$, green) excitonic states. For electric fields $|F| >$ 90~kV/cm electron tunneling is fast and the neutral charge states can be observed.}    
	\label{fig:Figure3}
\end{figure}
To further investigate the electron tunneling escape during the reset pulse we performed a series of PL measurements recorded as a function of the reset voltage V$_{reset}$ (fig~3a) whilst keeping V$_{readout}$ fixed at 0.8~V ($|$F$_{readout}|$=7~kV/cm) in the luminescence regime. Again the laser was turned OFF during the reset pulse, such that dot charging and luminescence occurs during the readout phase of the measurement. The results of these measurements are presented in fig~3. Figure~3b shows luminescence spectra recorded with reset fields $|$F$_{reset}|$ ranging from 14 to 170~kV/cm. In fig~3c the combined emission intensity of the uncharged (X$^0$ + 2X$^0$), singly charged (X$^-$+ 2X$^-$) and doubly charged (X$^{2-}$) transitions is plotted as a fraction of the total luminescence yield. We note that this total intensity remains constant over the entire measurement range, demonstrating that all relevant transitions were considered in our analysis.
The neutral charge states, shown in fig~3c, dominate in the high field region ($|$F$_{reset}| >$ 90~kV/cm) , whilst the doubly charged exciton is strongest in the low field region ($|$F$_{reset}| <$ 60~kV/cm). At the transition ($|$F$_{reset}|$ = 78~kV/cm) the singly charged states (X$^-$ and 2X$^-$ in fig~3c) show a maximum in their intensity. The observed behavior is consistent with electric field dependent electron tunneling during the reset pulse as expected. We observe emission from charged excitonic states only over a field range where the electron tunneling time is longer than the 400~ns duration of the applied reset pulse. It is only in this region that charge accumulation can occur over several measurement cycles. On the other hand when the electron tunneling time is shortened, due to the application of high electric fields, the optically induced charge will be removed by each reset pulse and emission from the neutral charge states are restored. Because of the interparticle Coulomb repulsion, a doubly charged dot is more likely to loose a charge at a fixed value of electric field, leaving the dot in a singly charged configuration. This simple consideration accounts for the earlier onset of the intensity decrease for the X$^{2-}$ compared to the singly charged states observed in fig~3c and leads to an intermediate increase of the intensity of those states with its maximum at $|$F$_{reset}|$ = 78~kV/cm. The saturation value of 20\%, in the high field region for X$^-$, is attributed to charging from the back contact that becomes much faster than the optical charging rate discussed above.
\\
\begin{figure*}[h]
	\centering
		\includegraphics[width=1.0\textwidth]{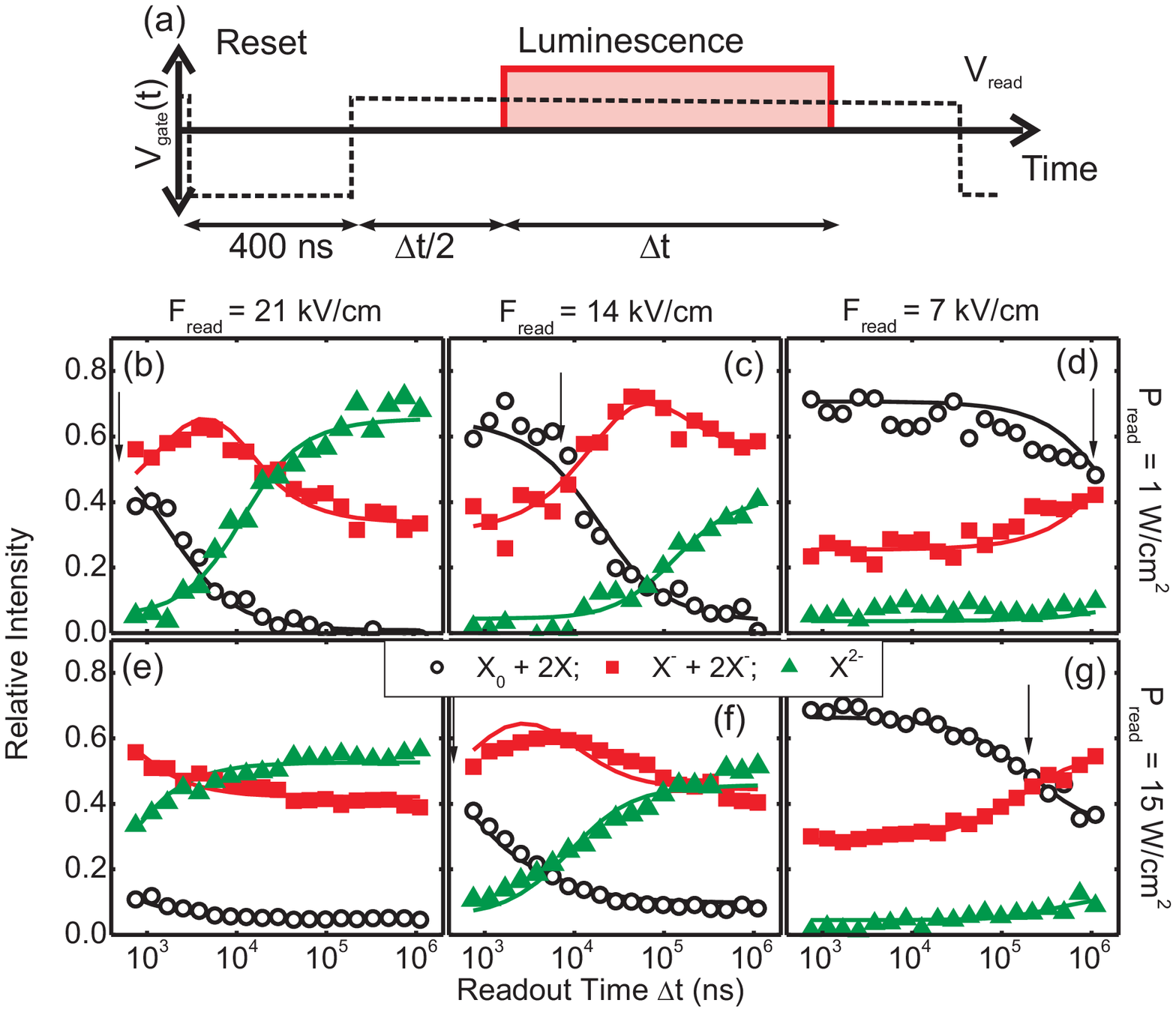}
	\caption{(Color online) (a) Schematic representation of the pulse sequence, $\Delta$t is varied during the experiment. (b-g) Time evolution of the relative intensities of the charge neutral (X$^0$+2X$^0$, black), singly charged (X$^-$+2X$^-$, red) and the doubly charged (X$^{2-}$, green) excitonic states. The rows show measurements with constant illumination power (b-d) P$_{readout}$= 1~W/cm$^2$ (e-g) P$_{readout}$= 15~W/cm$^2$;whereas columns depict constant electric field (b,e) F$_{readout}$=21~kV/cm, (c,f) F$_{readout}$=14~kV/cm, (d,g) F$_{readout}$=7~kV/cm.}
	\label{fig:Figure4}
\end{figure*}
Up to now we have shown that the device can be used to optically charge the QD under certain conditions and reset the dot to a neutral state. To detect the charge state of the dot in the same structure via the energy shift of the charged excitonic lines \cite{Warburton00, Finley01}, it must be possible to switch off optical charging by controlling the gate potential during readout. To further extract quantitative information about the charging dynamics we continue to discuss measurements of the relative intensities of the excitonic states as a function of the readout time, excitation power and F$_{readout}$. The results of these measurements are presented in the various panels of fig~4. To ensure that the QD is empty before illumination begins a reset pulse is applied periodically before each illumination cycle as discussed above. The pulse sequence used for these experiments is schematically depicted on in fig~4a. Following a 400ns reset pulse the gate is held at V$_{readout}$ and the laser is turned ON for a time $\Delta$t$_{read}$. The illumination is well separated from the voltage transition by a delay of $\Delta$t$_{read}/2$. The upper row of panels in fig~4b-d shows measurements performed for V$_{readout}$ such that the static electric field is 21~kV/cm, 14~kV/cm and 7~kV/cm at low optical excitation power of 1~W/cm$^2$. Similarly, the lower row of panels (fig~4d-f) show the same data recorded at a much higher readout power P$_{readout}$ = 15~W/cm$^2$. The readout time $\Delta t_{read}$ is plotted on a logarithmic scale and spans the range from 700~ns to 1~ms. For a clearer representation, the sum of the intensities of the charge neutral states (X$^0$ + 2X$^0$) is presented in fig~4 by the black open circles, the sum of the intensities of the singly charged states (X$^-$ + 2X$^-$) is shown by red filled squares and the doubly charged (X$^{2-}$) exciton emission is plotted by green filled triangles. We begin by focusing our discussion on fig~4c, in the center of the upper row of panels (fig~4b-d). For small $\Delta$t$_{read}$, the uncharged states X$^0$+2X$^0$ dominate, but decrease in intensity with increasing illumination time, due to optical charging. As expected, the singly charged states (X$^-$ + 2X$^-$) exhibit an anticorrelated behavior, when compared to (X$^0$ + 2X$^0$) over the first 100~$\mu$s. In the range from $\Delta$t$_{read}$ = 100~$\mu$s to 1~ms the signal from singly charged states again decreases, whilst the doubly charged states gain intensity. This illustrates nicely the sequential optical charging dynamics in the QD monitored for this excitation power and electric field in real time. For short illumination times little charge builds up and the X$^0$+2X$^0$ intensity dominates as seen above. As $\Delta$t$_{read}$ increases, there is more time to charge up the dot with electrons and, consequently, the average electron occupancy in the dot increases. At $\Delta$t$_{read}$=64~$\mu$s we find that a single electron occupies the QD on average. If $\Delta$t$_{read}$ increases further, a second electron is added to the dot , leading to an increase of the doubly charged states, and an anticorrelated reduction of singly charged transitions. 
\\
Since optical charging is expected to become more efficient as the hole tunneling becomes faster, these characteristics should shift towards shorter timescales as we increase the electric field during the illumination time. Precisely this behavior can be seen in fig~4c for a higher electric field of $|$F$_{readout}|$=14~kV/cm. In contrast to $|$F$_{readout}|$=21~kV/cm, the charging dynamics of the QD slows down as the electric field reduces and the device is operated closer to flat band as shown by fig~4d.\\
Naturally, an enhanced charging rate is also expected when the optical power is increased during the readout cycle. This expectation is confirmed by our measurements at higher power. These results are presented in figs~4d-f, that reveal a clear shift of the evolution of neutral, singly and doubly charged transitions towards shorter times when the readout power density is increased to 15~W/cm$^2$, compared to 1~W/cm$^2$.
\\
A 3-level rate equation can be used to quantitatively describe this data and extract the 1e and 2e optical charging rates ($\Gamma_{ch1}$ and $\Gamma_{ch2}$) and their dependence on optical power P$_{readout}$ and static electric field F$_{readout}$. The combined intensity of the X$^0$ and 2X$^0$ lines are a measure of the fraction of the time when the dot is uncharged. Similarly, the combined intensity of singly (X$^-$ + 2X$^-$) and doubly X$^{2-}$ charged state reflect the probability to find a dot with 1e or 2e respectively. To simplify our analysis, we assume that electrons and holes are trapped geminately into the QD such that single holes cannot be captured and charging of the dot is always governed by the hole tunneling time. This is a fairly good assumption, since we excite at the low energy edge of the wetting layer and excitons are formed before capture into the dot.
Within this framework we set up a three level system as depicted schematically in the inset of fig~5. The three levels N$_0$, N$_1$, N$_2$ represent generically the uncharged, singly charged and doubly charged dot. Higher charged levels are not taken into account, as no indication for such higher order charged states was observed in the measurements. We assume different charging rates for the first ($\Gamma_{ch1}$) and second ($\Gamma_{ch2}$) electron since the hole tunneling time increases when tunneling from a negatively charged dot due to the attractive Coulomb interaction between the particles trapped in a dot. Since we are interested in timescales up to 1~ms, we also include generic charge loss rates $\Gamma_{loss1}$ and $\Gamma_{loss2}$ for the first and the second electron respectively. In the framework of this model the time dependent rate equations describing the charging dynamics have the form:
\begin{eqnarray}
	\frac{\partial N_0(t)}{\partial t} = - N_0(t)\cdot \Gamma_{ch1} + N_1(t) \cdot \Gamma_{loss1} \\
	\frac{\partial N_1(t)}{\partial t} = N_0(t)\cdot \Gamma_{ch1}
	+ N_1(t) \cdot (\Gamma_{ch2}-\Gamma_{loss1}) + N_2(t)\cdot \Gamma_{loss2} \\
	\frac{\partial N_2(t)}{\partial t} = N_1(t)\cdot \Gamma_{ch2} - N_2(t) \cdot \Gamma_{loss2} \\
\end{eqnarray}
To ensure that the functions N$_{0,1,2}(t)$ reflect charge occupancy probabilities, we normalize them such that N$_0$(t)+N$_1$(t)+N$_2$(t)=1 for all t. Since the luminescence is integrated over the readout time $\Delta$t$_{read}$, time, the fit function has the form $I(t)= \frac{1}{\Delta t_{read}} \int N_n(t) dt$, where n $\in$ [0,1,2] and the integration is performed from 0 to $\Delta t_{read}$. The model was fitted to the measurements using a genetic algorithm \cite{Mitchell96} and the standard error was calculated for these fits \cite{Press86}. As can be seen from fig~4 this simple model fits the data quite well using a single set of fit parameters ($\Gamma_{ch1}$, $\Gamma_{ch2}$, $\Gamma_{loss1}$, $\Gamma_{loss2}$) for the time evolution of all three charge states. The extracted charging rates are summarized in fig~5a ($\Gamma_{ch1}$) and fig~5b ($\Gamma_{ch2}$) as a function of readout power and electric field. As expected, both $\Gamma_{ch1}$ and $\Gamma_{ch2}$ increase linearly with power for all measured electric fields. Hence, for each value of the applied electric field we can calculate the 1e and 2e charging rates normalized to the incident power P$_{readout}$. The result of this analysis is presented in fig~5c. These results clearly demonstrate that charging is significally more effective for the first optically generated electron than the second due to the attractive Coulomb interaction. We attribute this finding to the suppressed tunneling probability of a hole from the dot when two electrons are present. Furthermore, the decrease of the normalized charging rates varies exponentially with the electric field as excepted reflecting the field dependent tunneling probability of the hole. In contrast to the charging rates, the fitted loss rates $\Gamma_{loss1}$ and $\Gamma_{loss2}$ are typically $\approx$10x smaller. An exponential dependence on electric field and linear increase of the loss rates with illumination power suggest that the charge loss arises both due to tunneling escape of the electrons through the AlGaAs barrier and also photo induced discharging \cite{Heinrich01}. Nevertheless, independent measurements on the charge stability in the absence of illumination revealed that charge loss due to tunneling takes place over timescales much longer than 10~$\mu$s. This clearly demonstrates that optically induced charge storage is possible over much larger timescales, controlled by the gate potential used. Moreover for charge readout the charging rates $\Gamma_{ch1}$ and $\Gamma_{ch2}$, although small, are identified as the fastest process limiting long readout times. To illustrate the implications of these results we calculated the fidelity of a ``charge occupancy" measurement. Such a measurement could be conducted, for example, by switching the device to a regime where $\Gamma_{ch1}$ and $\Gamma_{ch2}$ are very low, and using the relative PL yield of X$^-$ and X$^{2-}$ to probe the charge status of the dot. To estimate the fidelity we set a criteria for the charge occupancy; if X$^{2-}$ is the dominant emission line then we declare the dot occupancy to be 2e, whereas if X$^-$ dominates it is declared to be 1e. If we fix $\Gamma_{loss_2} = \Gamma_{ch2}$ = 1000~s$^{-1}$, values easily achievable for low illumination powers and low electric fields as shown above, and set a readout time of 100~$\mu$s, we calculate from our simple model a fidelity of over 95\% to correctly deduce the 1e or the 2e charge state. This fidelity can be increased further by choosing either a lower excitation power or a shorter readout time $\Delta t_{read}$. Our experimental results indicate that electrical charging of the dots from the back contact is not negligible, since even for very short illumination $\Delta t_{read} <$1~$\mu$s and low power the fraction of emission from singly (doubly) charged states can not be reduced below 20\% (5\%). Therefore, initial values N$_1$(0)=0.2 (N$_2$(0)=0.05) were used in the model. In future experiments charging from the back contact could be reduced by increasing the separation of the QDs from the n-doped layer.   
\\
\begin{figure}[h]
	\centering
		\includegraphics[width=1.0\textwidth]{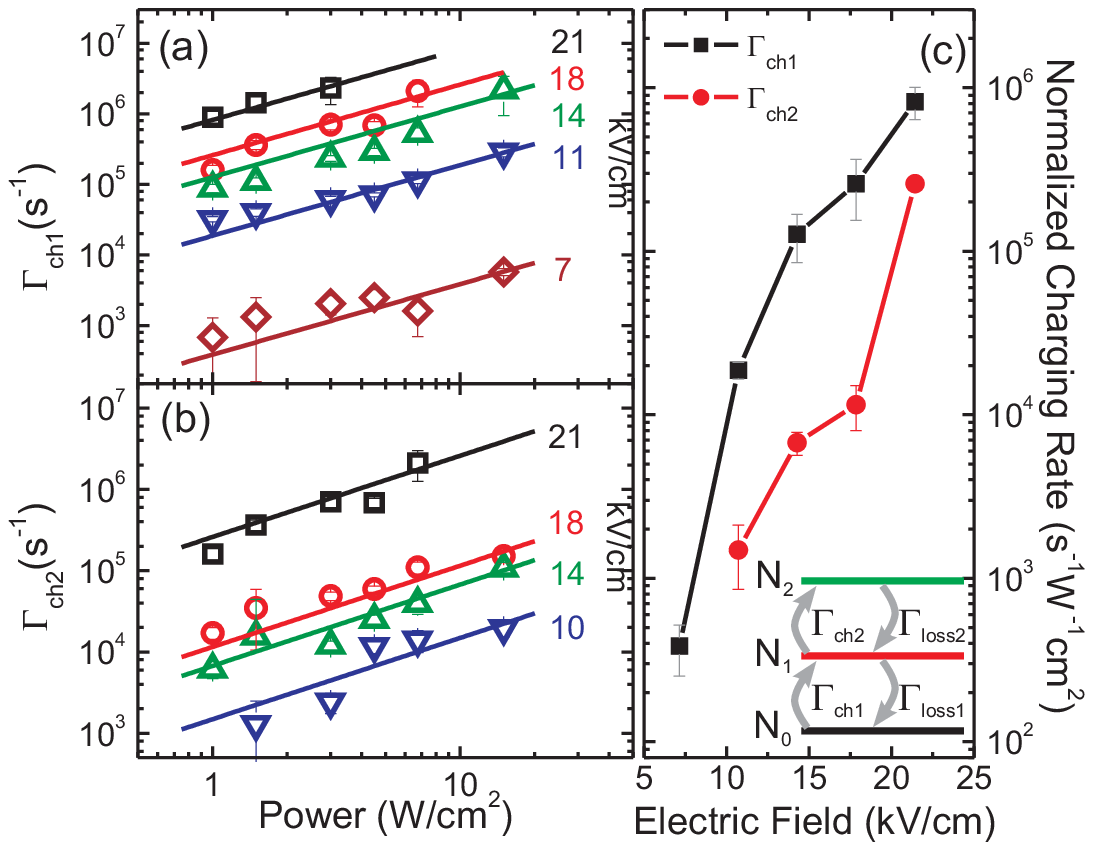}
	\caption{(Color online) Charging rates $\Gamma_{ch1}$ (a) and $\Gamma_{ch2}$ (b) as a function of power density for different applied electric fields. The charging rates increase linearly with incident power for all electric fields observed.(c) Charging rates $\Gamma_{ch1}$ (black, squares) and $\Gamma_{ch2}$ (red, circles) as function of applied field normalized for incident power density. The rates vary exponentially on the applied electric field since the charging mechanism is governed by hole tunneling. The inset shows the three levels used N$_0$, N$_1$, N$_2$ in our rate equation analysis of the optical charging.  }
	\label{fig:Figure5}
\end{figure}
Let us summarise the results from the time resolved data shown in fig~4: firstly, the structure can be biased in a regime of high electric fields, where charging is very effective with charging rates $\Gamma_{ch1} \gg $ 10$^{6} s^{-1}W^{-1}cm^2$. Moreover, the structure can be switched into a readout regime with a negligible charging rate $\Gamma_{ch2} < $ 1000 $s^{-1}W^{-1}cm^2$ and even lower loss rates. When combined, these properties allow high fidelity charge measurements.
\\
\begin{figure}[h]
	\centering
		\includegraphics[width=1.0\textwidth]{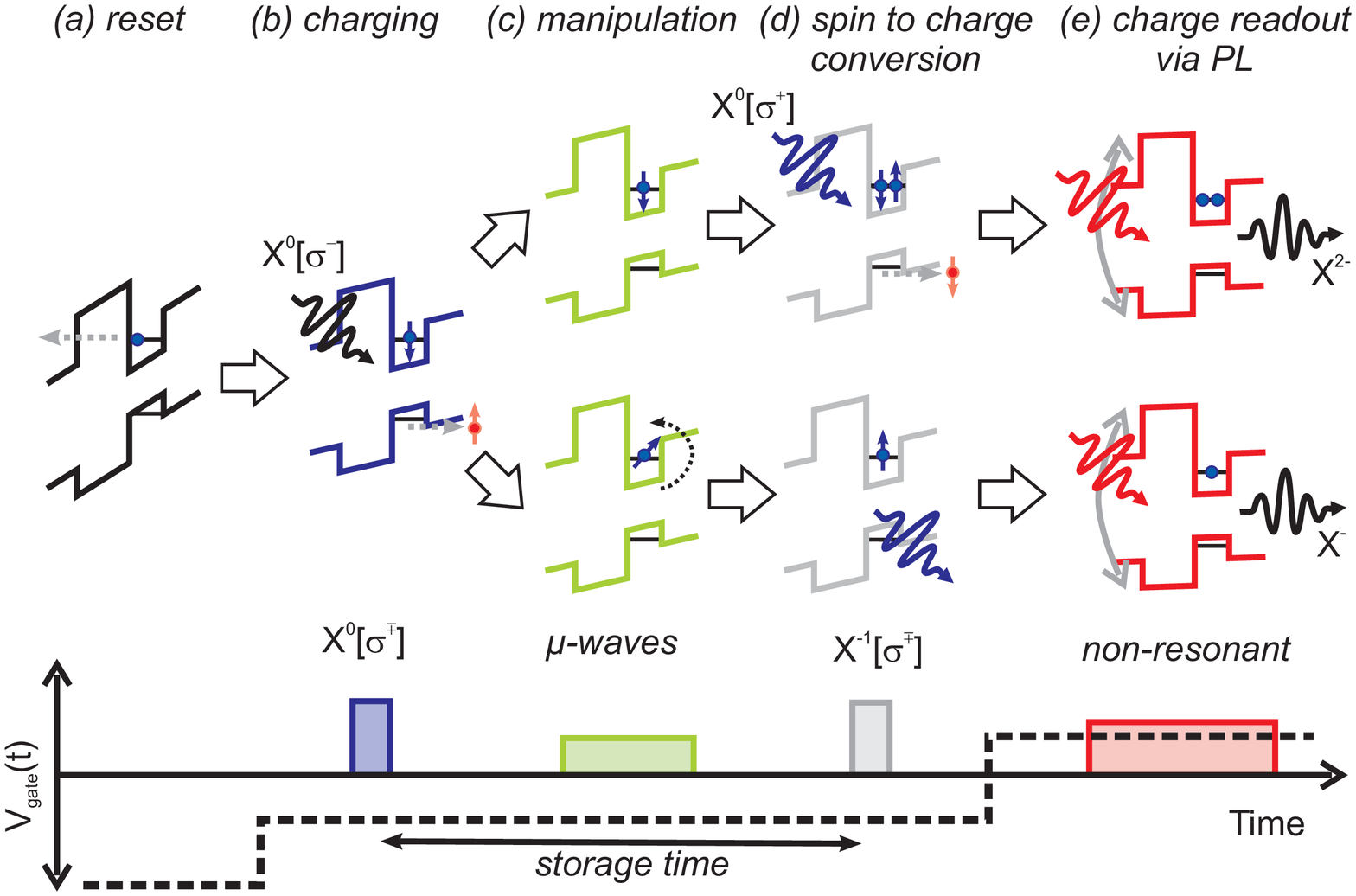}
	\caption{(Color online) Schematic representation of the spin readout scheme. The individual steps are: (a) reset by fast electron tunneling, (b) charging with resonant excitation followed by hole tunneling, (c) spin manipulation with e.g. microwave pulses, (d) spin to charge conversion by spin conditional resonant absorption, (e) charge readout via non-resonant photoluminescence. }
	\label{fig:Figure6}
\end{figure}
Our results have demonstrated that reliable charge measurements can be performed on a single QD. We proceed by proposing a spin readout mechanism for single dots based on this concept. Figure~6 illustrates schematically a scheme for spin generation, manipulation, initialization and readout.  
Each measurement cycle starts by resetting the QD to an empty state to avoid cumulative errors. This is done simply by applying high electric fields that ensure tunneling escape of the electrons (fig~6a). Following this, the field is reduced to switch the device into charging mode and a single electron with a defined spin orientation is optically pumped into the QD illuminating the device with laser light tuned to the X$^0$ transition. Using circularly polarized light the spin orientation of the optically generated electron is controlled by the helicity of the polarization \cite{Bayer02} (fig~6b) and for the purposes of illustration, in fig~6 we depict the creation of a spin down electron using $\sigma^-$ polarized light. Following initialization, the spin can be stored, thermally relax or be externally manipulated for example using microwave pulses for a time $\Delta t_{store}$. We note that during the storage time the sample is not illuminated, and the static electric field experienced by the QD in the intrinsic region of the diode can be varied over a wide range. In fig~6c we schematically depict a $\pi$ - rotation of the spin in the lower row, whilst the spin in the upper row of images remains unchanged. After such controlled manipulation, the spin orientation is measured by applying a second circularly polarized laser pulse tuned to the X$^-$ transition. A laser pulse at X$^-$ with $\sigma^+$ helicity would create an additional spin down electron if the spin orientation of the resident electron is up, while the charge status would remain unperturbed if the resident electron is spin down. This step performs a spin to charge conversion; after the optical pulse tuned to X$^-$ one or two electrons populate the dot depending on the spin projection of the resident electron. In a final step in the measurement, the device is biased away from the optical charging regime and time integrated PL is used to non pertubatively test the charge occupancy (1e or 2e) and, thus the spin orientation of the resident electron. As discussed above, the charge state of the dot can be deduced with high fidelity from the relative intensity of the X$^-$ or the X$^{2-}$ luminescence lines. Our results show, that PL-readout only weakly disturbes the charge state of the dot for timescales up to 100~$\mu$s, and, therefore, the storage time can be increased accordingly. 
\\
In summary, we have demonstrated a device that allows controlled optical charging of a single QD with an electron, read out of the charge state by time integrated PL and controlled discharging to reset the sample and empty the dot. For readout, we showed that further charging of the dot can be strongly suppressed under flat band conditions and weak illumination P$_{read}$=1~W/cm$^2$ allowing reliable determination of the charge state by comparing the intensity of the X$^-$ and X$^{2-}$ charged exciton emission lines. Finally, we suggested an all optical spin readout concept that is based on spin to charge conversion using such devices. With this concept it should be possible to optically initialize a spin, manipulate it on millisecond timescales and reliably read the spin state with sufficient signal, combined properties that are not offered by other optical spin readout methods.
\\
The authors gratefully acknowledge financial support by the \textit{DFG} via \textit{SFB}-631 and the German Excellence Initiative via the \textit{Nanosystems Initiative Munich} (NIM).

\end{document}